\documentclass[final]{aipproc}%
\usepackage[english]{babel}
\usepackage{amsmath}
\usepackage{amsfonts}
\usepackage{amssymb}
\usepackage{graphicx}%
\layoutstyle{8x11double}
\begin{document}

\title{Moments with Yakov Borisovich Zeldovich}


\author{Remo Ruffini\footnote{to appear in the proceedings of the international conference ``The Sun, the Stars, The Universe and General Relativity'' in honor of Ya.~B.~Zeldovich 95$^{\rm th}$ Anniversary, held in Minsk, Belarus on April 20-23, 2009.}}
{address={ICRANet, p.le della Repubblica, 10 - 65122 Pescara, Italy and \\ ICRA and University of Rome ``Sapienza'', p. Aldo Moro 5, I-00185, Rome, Italy and \\ ICRANet, University of Nice-Sophia Antipolis, 28 avenue de Valrose, 06103 Nice Cedex 2, France} }

\begin{abstract}
A recollection of special moments spent with Yakov Borisovich Zeldovich and with the scientists of Soviet Union and abroad.
\end{abstract}

\maketitle

The first impression upon %bob 
meeting a person is the one which characterizes 
all subsequent %bob the entire further 
interactions.

I met Yakov Borisovich Zeldovich for the first time in 1968 at the GR5 meeting in Tbilisi. I had known his name from his two classic papers on relativistic astrophysics in Physics Uspekhi coauthored with Igor Novikov \cite{ZN1,ZN2}. 
%bob
There had been a strong impulse to boycott the GR5 meeting due to the tense relations over human rights between 
the %bob
Soviet Union and the USA at that time. Finally a small group around Johnny Wheeler decided to participate. Among them were Arthur Komar, Bruce Partridge, Abe %bob
Taub and myself. 

It was also my first visit to 
the %bob
Soviet Union. The entrance 
to %bob in 
Leningrad was already very special showing the difference 
in %bob of 
organization from our Western world. I will recall elsewhere some of the anecdotes. It was in the airplane to Tbilisi that a very particular experience occurred. The year %bob
1968 was a time in which dissent was growing in the Soviet Union and the New York Times had just written an article on Andrei Sakharov and his reflections on peaceful coexistence and intellectual freedom. I boarded %bob 
the plane for Tbilisi with Arthur Komar. We sat in %bob on 
the last row of a quite modern jet  plane with open seats and shining windows, and we were commenting and laughing on all those stories we had heard in the West about windowless seats %bob
reserved for westerners  %bob
on Soviet planes. When the plane was almost full the stewardess called the names of  Arthur Komar and Remo Ruffini asked us to move to  %bob
seats reserved for us in the front of the plane. We were delighted and we considered this an honor. Our two seats were in a line of three seats $\ldots$ the only ones in the plane without a %bob
window. We were quite upset. In between us there was a third person who did not seem to speak English. So we started complaining about these methods and commenting appropriately also about Sakharov's %bob 
recent opinions as presented in the New York Times and asking ourselves about the fate of Sakharov after his open statements. The plane was supposed to be a direct flight to Tbilisi of approximately seven hours. After approximately three hours of flight, without any announcement, the plane abruptly started to descend quite rapidly and landed in a town called Mineralnye Vody. After landing there was a lot of confusion, there were additional planes and finally it was disclosed that, as a common practice in the Soviet Union in the %bob
presence of bad weather, the plane had stopped and we would %bob will 
continue the flight the morning after. It was also announced that for foreigners there would be {\emph a} room to sleep. Soon after %bob Just later 
I realized that there was only one room for all the foreigners! Since it was impossible to sleep I went back to the airport hall and I noticed this person who had been sitting %bob setting 
between me and Komar on the plane to be alone in the hall and had %bob having 
found a chair. He was seating quietly waiting for the morning. I was attracted by his silence and his self-control. I approached him introducing myself: ``Ruffini, Italy.'' To this his answer: ``Sakharov, Soviet Union!'' I still remember his serene smile. He was the first Soviet scientist %bob 
I met %bob meet 
on the way to our meeting in Tbilisi. The arrival in Tbilisi with Kumar and Sakharov was marked %bob signed 
by the fortunate encounter with other 
monumental scientific figures. %bob scientific monumental figures.

We had the marvelous opportunity to meet some historical figures like Vladimir Fock, Iosif Shklovsky and Alexei Petrov and also Dmitry Ivanenko. It was amusing to see the ceremonial relations between Fock and Ivanenko. Fock, 
who as %bob has 
expected 
was %bob
always in the first row, %bob
had a conspicuous auditorial ``apparat." Every time Ivanenko was taking the floor to speak, Fock was disconnecting his ``apparat'' with a very explicit gesture. In addition of course there was Yakov Borisovich surrounded by a large number of then young collaborators including Gennady Bisnovatyi-Kogan, Valery Chechetkin, Viktor Shvartsman, Nikolay Shakura, Alexei Starobinsky, Rashid Sunyaev, Sergei Shandarin and others. Zeldovich was encouraging all his students to attack in their scientific presentations almost like a boxer ring trainer.
\begin{figure}[!ht]
	\centering
		\includegraphics[width=6in]{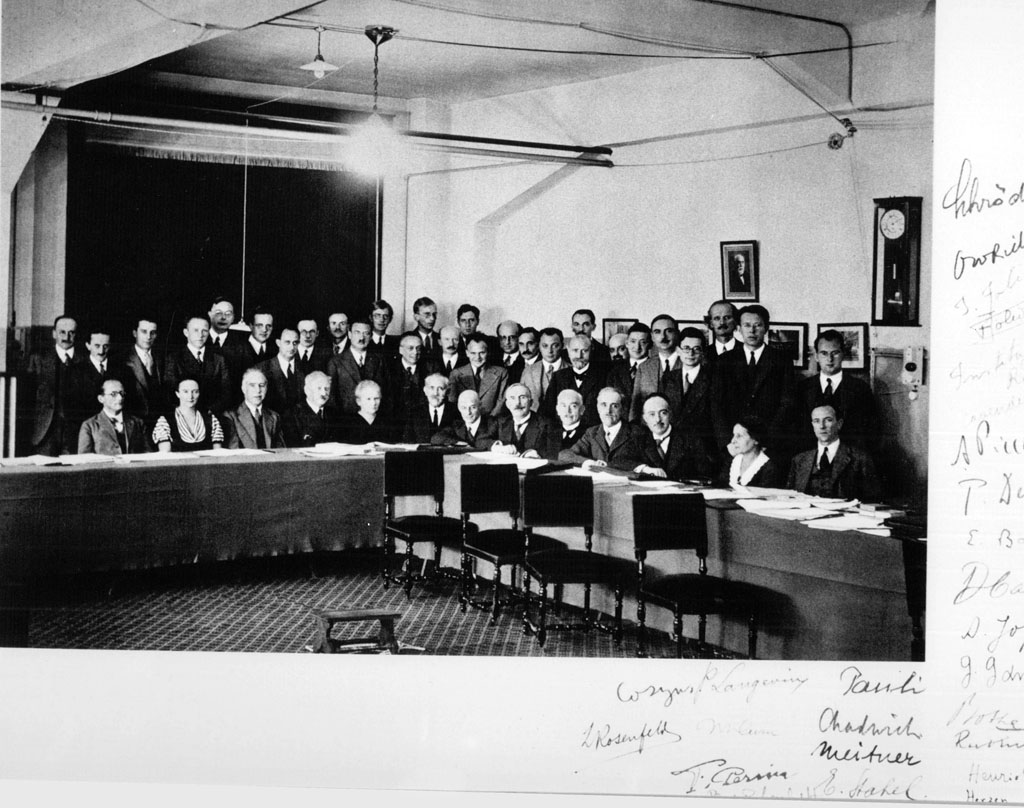}
	\label{fig:solvay1933}
	\caption{Solvay meeting of 1933. The series of photos from the Solvay meetings has been kindly given to ICRANet by Jacques Solvay, the descendant of Ernest Solvay in occasion of the assignment of the Marcel Grossman award to the Solvay foundation. Gamow is on the last row, perfectly symmetric with respect to other participants.}
\end{figure}

The first day of the meeting Zeldovich invited me to lunch and asked me just at the beginning to speak about my research. I started to explain my work on self-gravitating %bob
bosons I had started in Rome and just reconsidered after an interaction with 
the Pascual %bob Pascaul 
Jordan group in Hamburg. Indeed it was there that we realized that the previous treatment on Einstein-Klein-Gordon fields had a fatal error in the energy-momentum tensor leading to meaningless results. Later the correct work was completed by myself at Princeton and the published paper \cite{RB} became known as the paper in which the new concept of Boson Stars was introduced. %bob Boson Stars. 
After my %bob the 
first words Yakov Borisovich stopped me. I asked why. He stated ``How long did you speak?'' I answered ``approximately forty seconds.'' To that he replied ``If Landau would have been here he would have stopped you after twenty seconds.'' To that I immediately replied somewhat amused and self-confident ``I do not think so, I am sure Landau would have said how new is this idea and he would have approved my considerations.'' He followed then my presentation of the new results and more %bob A more 
polite and constructive discussions followed for the rest of the lunch. We also talked %bob discussed 
about George Gamow. Zeldovich recalled the animosity of all Soviet physicists towards %bob against 
Gamow since he did not return to Moscow after the famous Solvay meeting of 1933, see figure \ref{fig:solvay1933}. By this action Gamow hampered the possibility for %bob of 
all Soviet physicists to travel abroad after that date. He recalled %bob
how he was motivated by a matter of pure confrontation against Gamow for some time. As soon as Gamow presented the theory of a hot universe he himself presented an alternative theory of a cold universe, initially at zero temperature \cite{Zel62}. The process of building %bob 
up %bob of 
heavy elements was stopped in his theory by the presence of a degenerate sea neutrinos and only hydrogen would be born from an expanding Friedman universe. He stressed again, how %bob in 
building such a theory was motivated ideologically and politically. He recognized the crucial role of the Penzias and Wilson discovery of the cosmic microwave background radiation which disproved his `political' theory and proved instead the validity of Gamow's theory\footnote{I have made recollection of all this in a recent publication in \cite{Kerr}.}. He finally concluded ``Yes: although Gamow made many mistakes he is one of the greatest Soviet scientists!'' And then recalling the fundamental contributions Gamow made to the understanding of the DNA structure %bob s 
he asked: ``How many Nobel prizes did Gamow receive? Two?'' I answered: ``None.'' And I was surprised how distant he was from our world.

\begin{figure}[!ht]
	\centering
		\includegraphics[width=3in]{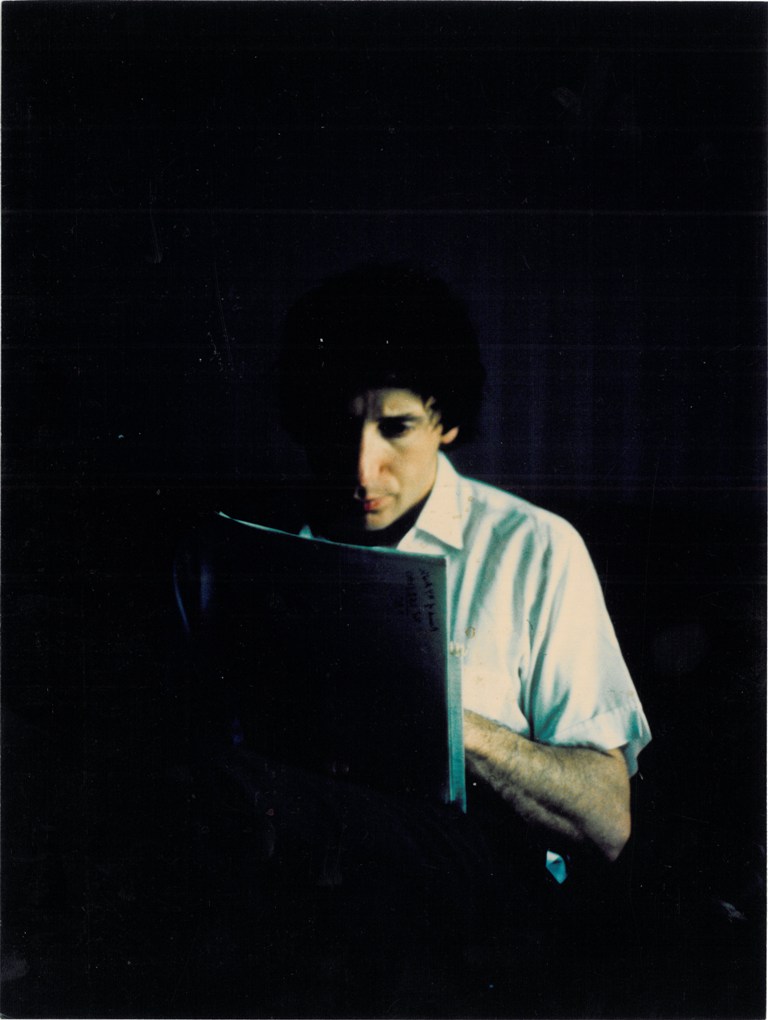}
	\caption{Picture of Viktor Shvartsman taken by myself in Moscow in 1975. Among the students of Zeldovich I was most impressed by Viktor. We reproduced one of his fundamental works in one of our book \cite{RRW}. It was clear to all of us that his isolation in the Caucasian mountains, so far from the world of Moscow and the world of theoretical research he was so strongly aiming for, was a key factor in the tragic epilogue of his life.}
	\label{Shvartsman}
\end{figure}

Paradoxically the work of neutrinos in cosmology was later reproposed by Viktor Shvartsman \cite{Shv69} by considering the role of the many neutrino species and in general to the number of ``difficult to observe particles with zero rest mass''. In that paper Viktor, see figure \ref{Shvartsman} established his classical result of an upper limit to the number of neutrino species $N_\nu\leq3$ assuming that the chemical potential of the electron neutrino be zero.
This result signed a new beginning in the dark matter problem in the Universe. I myself worked later on the role of massive neutrinos in cosmology. I considered their fundamental role both in cosmological nucleosynthesis \cite{BR} and in formation of the structure in the Universe due to dark matter, leading to a fractal structure of the Universe \cite{RST}.

\begin{figure}[!ht]
	\centering
		\includegraphics[width=3in]{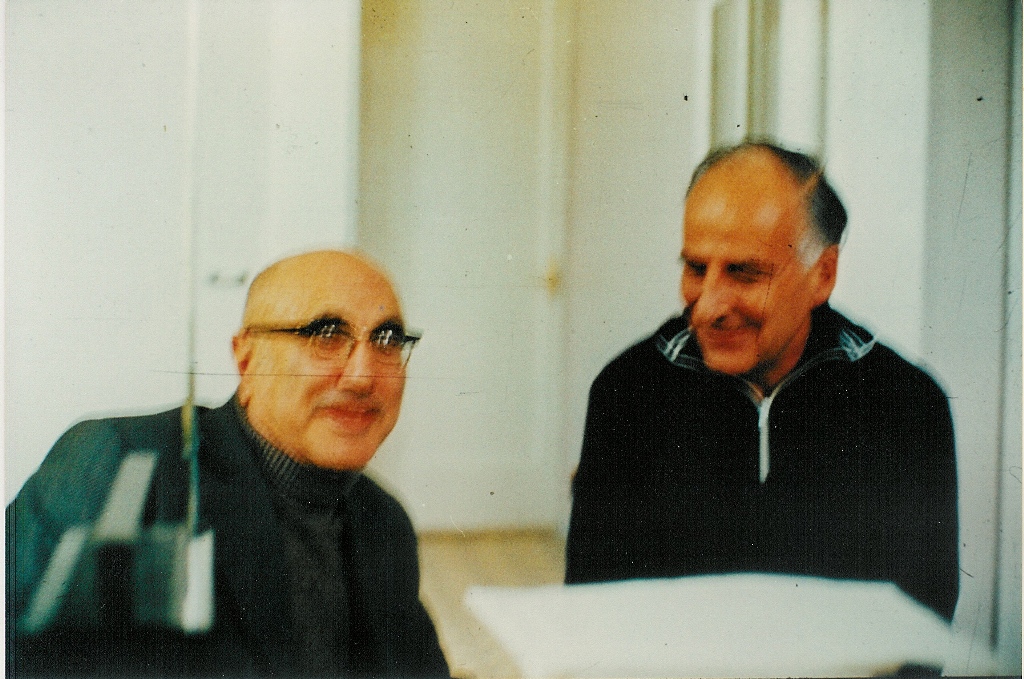}
	\label{fig:zp}
	\caption{Picture taken by myself in an unplanned visit to an hospital in Moscow. On the left side Zeldovich, on the right side Pontecorvo.}
\end{figure}

But let us go back to Zeldovich: we became very good friends in the following years, and I regularly met him in Moscow. We had also the great pleasure to share so many common friends. In particular, I remember many interactions with Bruno Pontecorvo, see figure \ref{fig:zp}.
In particular, with the participation of Bruno and Italian television we produced a documentary ``Il caso neutrino'' recovering the fundamental moments of the discovery of the neutrino all the way to the determination of their mass and their role in cosmology \cite{weblink}. 

Since 1973 I had the great fortune to become a very close friend of Evgeny Lifshitz. He had just granted to me and John Wheeler the honor of being quoted in a named exercise in the volume ``Theory of Fields'' of his classic 
series %bob treatise 
with Landau. As we became more familiar with Evgeny, I developed a profound admiration of his intellectual abilities, of his understanding of physics and of his moral stature. Evgeny often recalled a series of anecdotes. One of the best aphorisms of Landau: ``Astrophysicists often in error, never in doubt,'' and a different one related not only to astrophysicists but to physicists at large: ``Due to the shortness of our lives we cannot afford the luxury to spend time on topics which are not promising successful new results''. It was Evgeny who made me aware of some additional peculiarities in Zeldovich's character. 
\begin{figure}[!ht]
	\centering
		\includegraphics[width=3in]{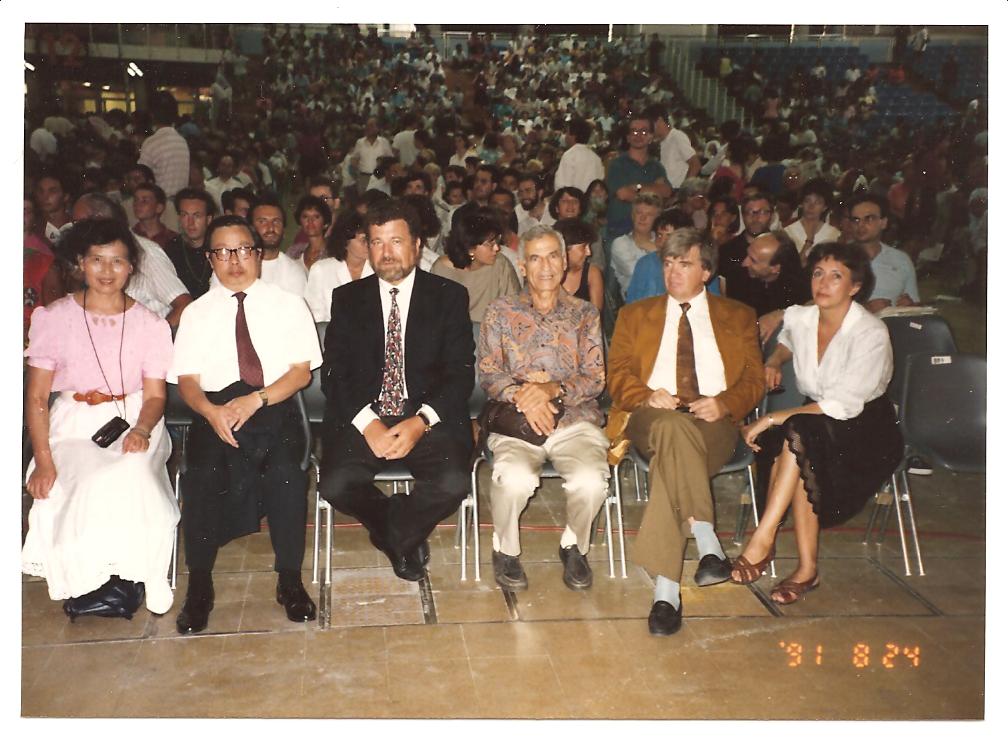}
	\caption{The picture of Li-Zhi Fang with his wife, myself, Leopold Halpern, Volodia Belinski and his wife at the Rimini Meeting of CL of 1991.}
	\label{fig:Belinski}
\end{figure}
\begin{figure}[!ht]
	\centering
		\includegraphics[width=3in]{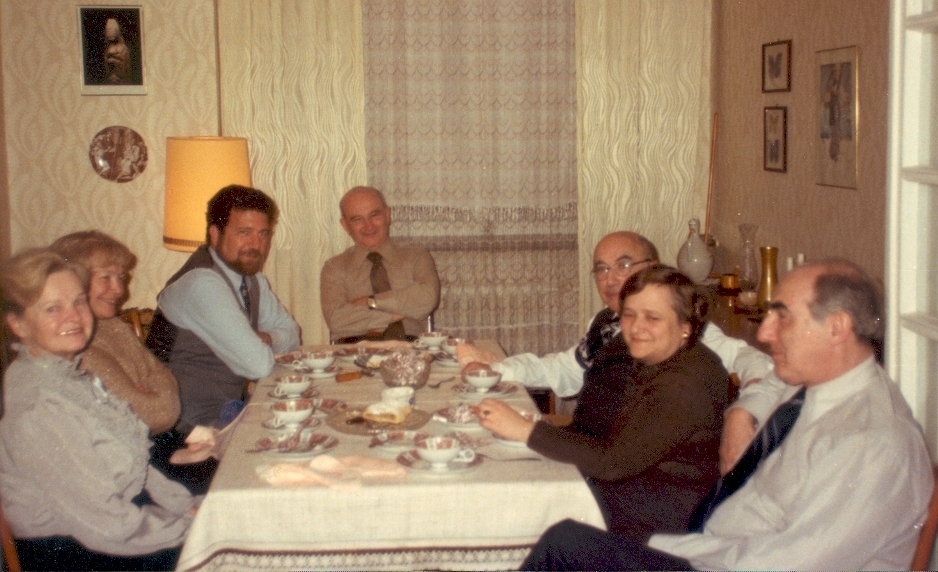}
	\label{rlzg}
	\caption{Dinner at Lifshitz home in Moscow (circa 1985). At the center Evgeny Lifshitz and, on his left, Zeldovich and Vitaly Ginzburg with their wifes. Picture taken by my wife Anna Imponente.}
\end{figure}
\begin{figure}[!ht]
	\centering
		\includegraphics[width=3in]{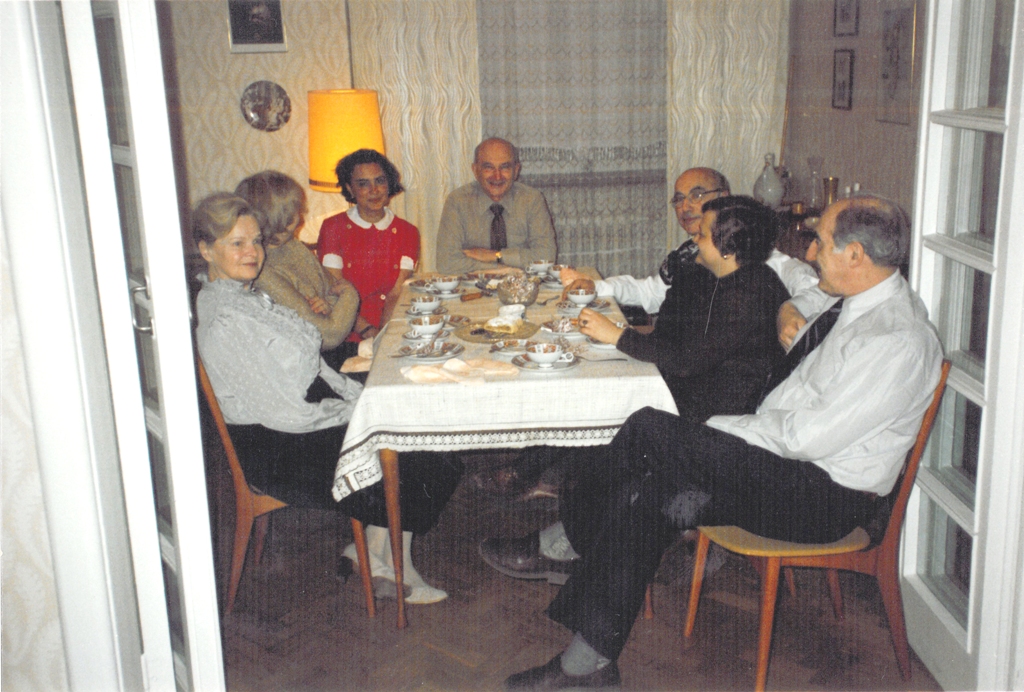}
	\label{rlzg2}
	\caption{Picture taken by myself.}
\end{figure}
\begin{figure}[!ht]
	\centering
		\includegraphics[width=3in]{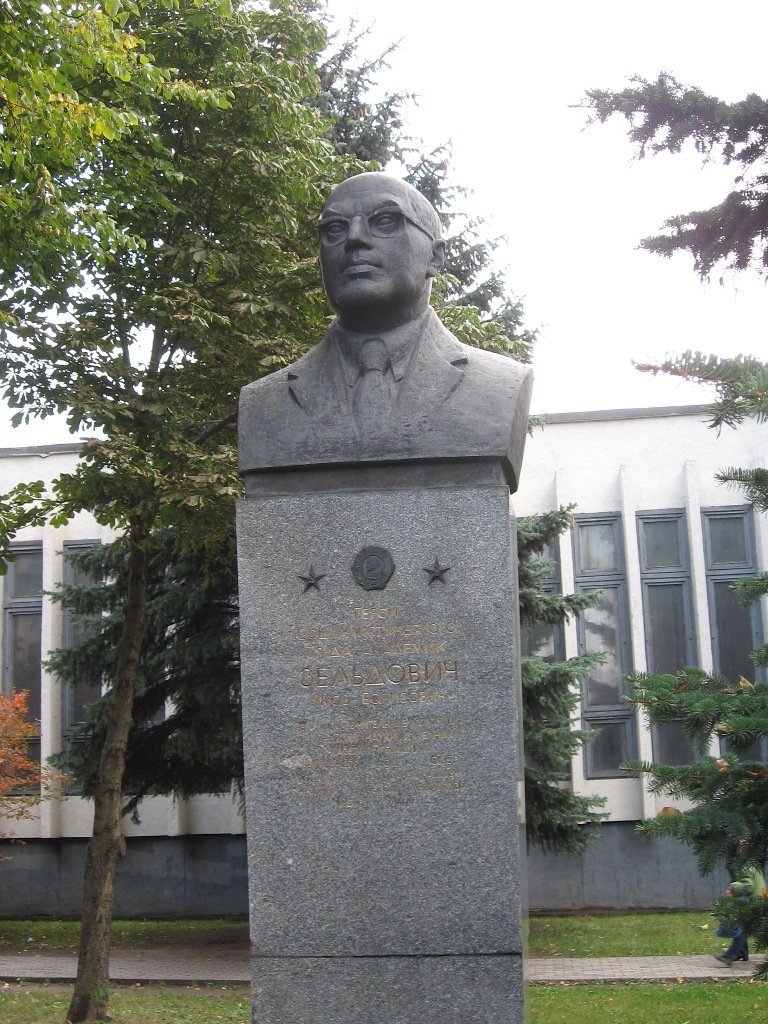}
	\label{m}
	\caption{Ya. B. Zeldovich monument in Minsk in front of National Academy of Sciences of Belarus.}
\end{figure}
\begin{figure}[ht]
	\centering
		\includegraphics[width=2.7in]{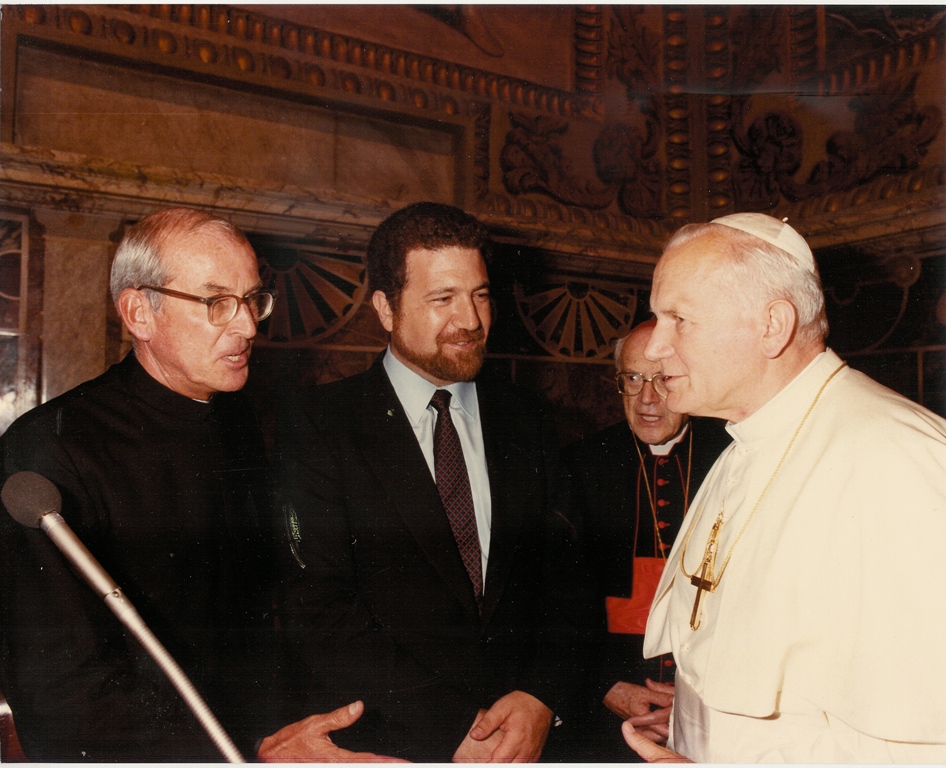}
	\caption{The picture of George Coyne and myself greeting John Paul II.}
	\label{fig:Coyne}
\end{figure}

Lifshitz described that famous argument %bob contention 
on the equation of state of neutron stars. Zeldovich first challenged the concept of the %bob
critical mass of the neutron star using %bob by 
an ad hoc model of supranuclear density interaction \cite{Zeldovich}. He had then purported the possibility of having an equation of state with the speed of sound equal to the speed of light, see \cite{Yak}. Lifshitz then recalled that %bob as 
Landau did not want ``to offend'' the intelligence of colleague physicists. If an issue was very difficult and important he would explain this issue. In other cases he was not going to explain and would ask %bobasking 
the person to answer himself. In the specific case of the extreme equation of state $p=\rho$ of Zeldovich he simply told him ``wrong!'', and to Zeldovich's request ``why?'' he simply answered ``you find out.'' This was before the tragic Landau car accident. After the accident Landau was no longer in any condition %bob not any longer in the condition 
to give a proof of the statement, and Zeldovich was unable to give a proof either. One day at the restaurant of the Academy in Leninsky Prospect, Yakov Borisovich asked Evgeny in my presence ``Why you did not insert my equation of state in the Landau and Lifshitz book?'' To this Lifshitz replied ``Did you solve the problem assigned by Landau?'', and to that Zeldovich said ``No.'', and to that Lifshitz's answer was ``Then I do not quote the result in the Landau and Lifshitz book.''
\begin{figure}[!ht]
	\centering
		\includegraphics[width=2.8in]{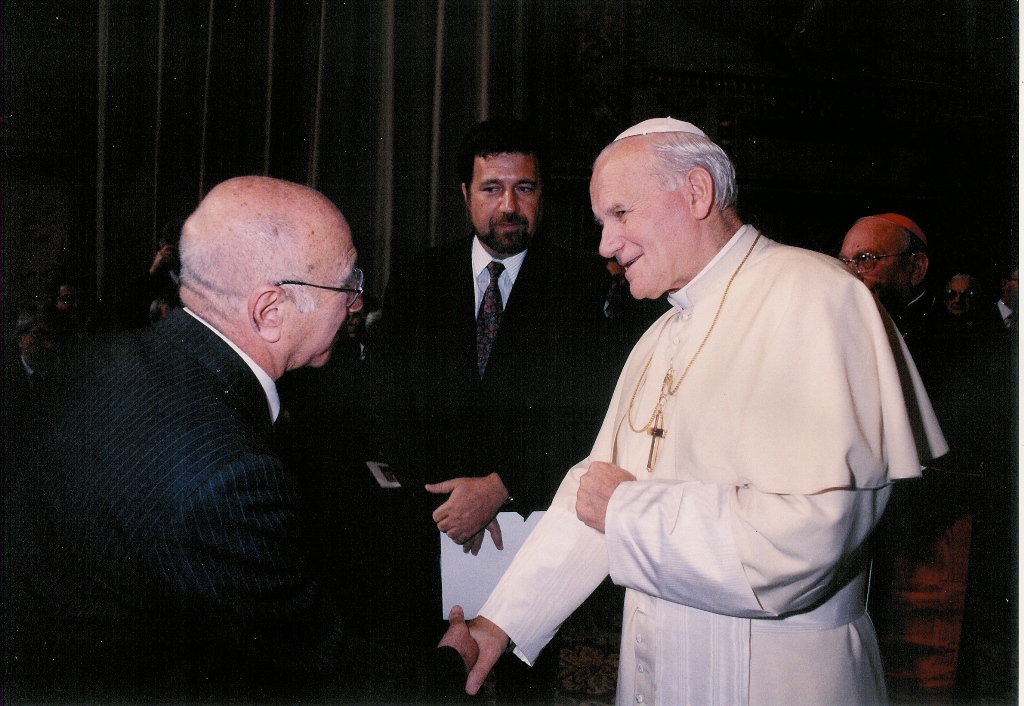}
	\label{p1}
	\caption{I look with terror Zeldovich approaching the Pope John Paul II
clearly with an unidentified object disguised under his jacket.}
\end{figure}
\begin{figure}[!ht]
	\centering
		\includegraphics[width=2.8in]{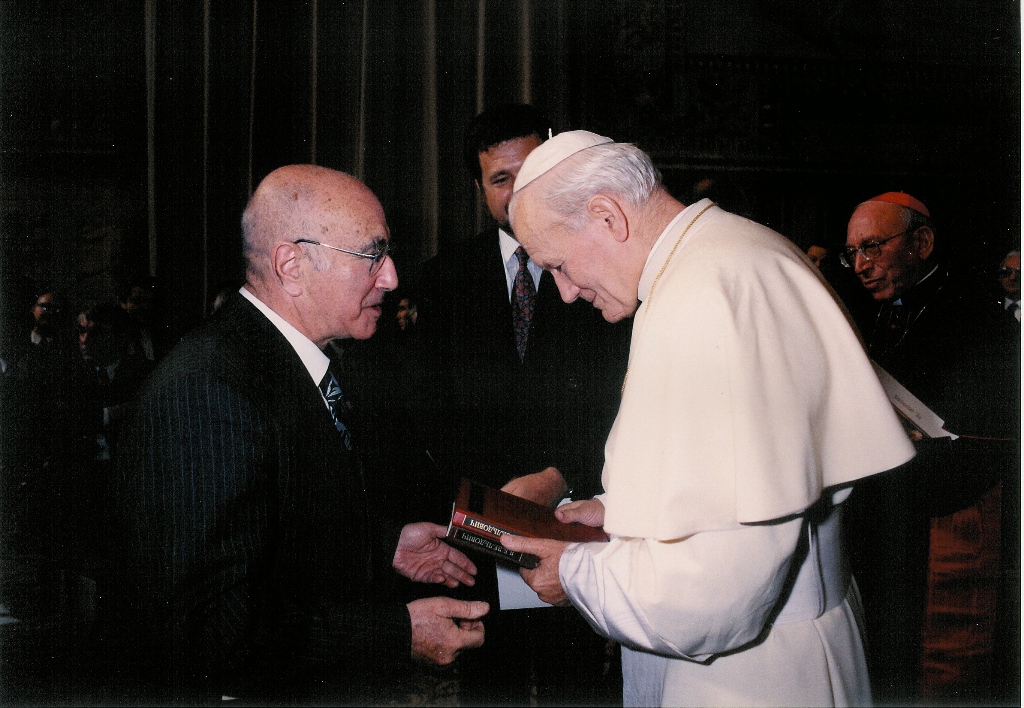}
	\label{p2}
	\caption{Zeldovich presenting his books to Pope John Paul II.}
\end{figure}
\begin{figure}[!ht]
	\centering
		\includegraphics[width=2.8in]{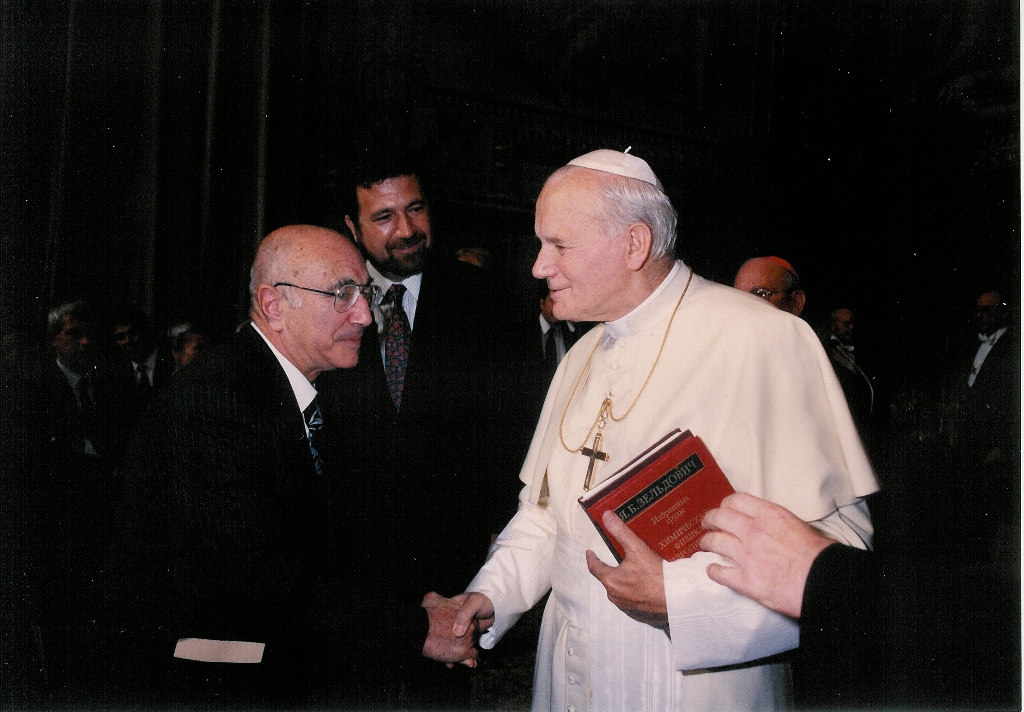}
	\label{p3}
	\caption{Zeldovich after the presentation of his books. To the offering of the books the Pope said ``Thanks'' and Zeldovich very loudly shouted ``Not just `thanks' ! These are fifty years of my work!'' The Pope kept Zeldovich's collected papers under his arm during the entire rest
of the audience.}
\end{figure}

My visit to Moscow was specially joyful due to the interactions with so many extraordinary scientists like Aleksandr Prokhorov, Isaac Khalatnikov, Pavel Cherenkov, Vitaly Ginzburg and others kindly invited to lunch with me in the Italian Embassy by the then Italian ambassador Sergio Romano and his predecessors. Encounter with Khalatnikov was especially productive. Khalat was the founder of the Landau Institute. However, among the others faculty members was Vladimir Belinski. The friendship with Lifshitz and Khalat soon extended to Volodia. So much so, that it transfered to Italy with his wife Elena, see figure \ref{fig:Belinski}, and became Italian citizen and one of the first faculty members of the newly founded ICRANet since 2005. Also extremely pleasant were the meetings at Yevgeny's home with friends and their wives, see figures \ref{rlzg} and \ref{rlzg2}.
\begin{figure}[!ht]
	\centering
		\includegraphics[width=2.7in]{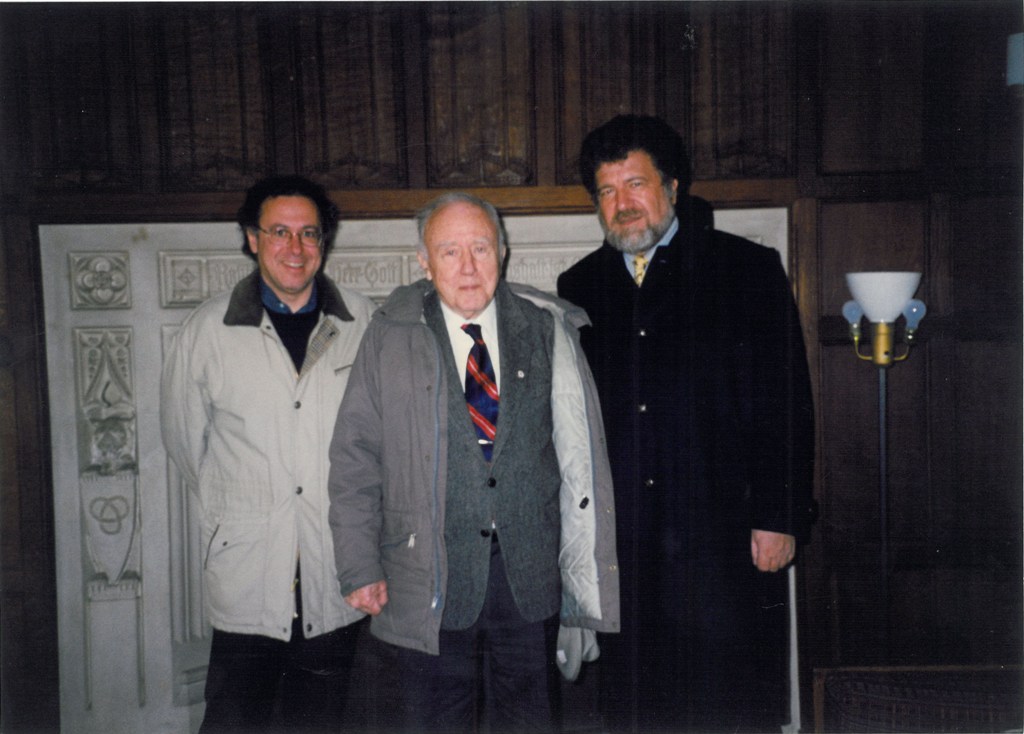}
	\caption{Picture of Wheeler, Christodoulou and myself in Fine Hall in Princeton in the former office of Albert Einstein. The picture is taken in front of the fireplace where Einstein wrote with charcoal, and now is engraved in gothic scripture in the marble, the famous sentence ``Raffiniert ist der HerrGott, aber boshaft ist er nicht''.}
	\label{fireplace}
\end{figure}
\begin{figure}[!ht]
	\centering
		\includegraphics[width=2.7in]{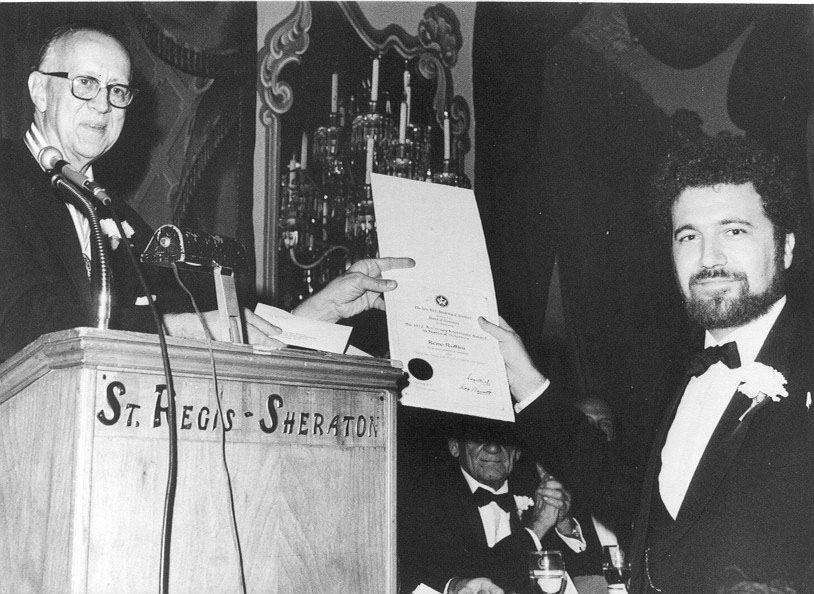}
	\label{prize}
	\caption{Receiving the Cressy Morrison Award of the New York Academy of Sciences in 1972.}
\end{figure}
\begin{figure}[!ht]
	\centering
		\includegraphics[width=6in]{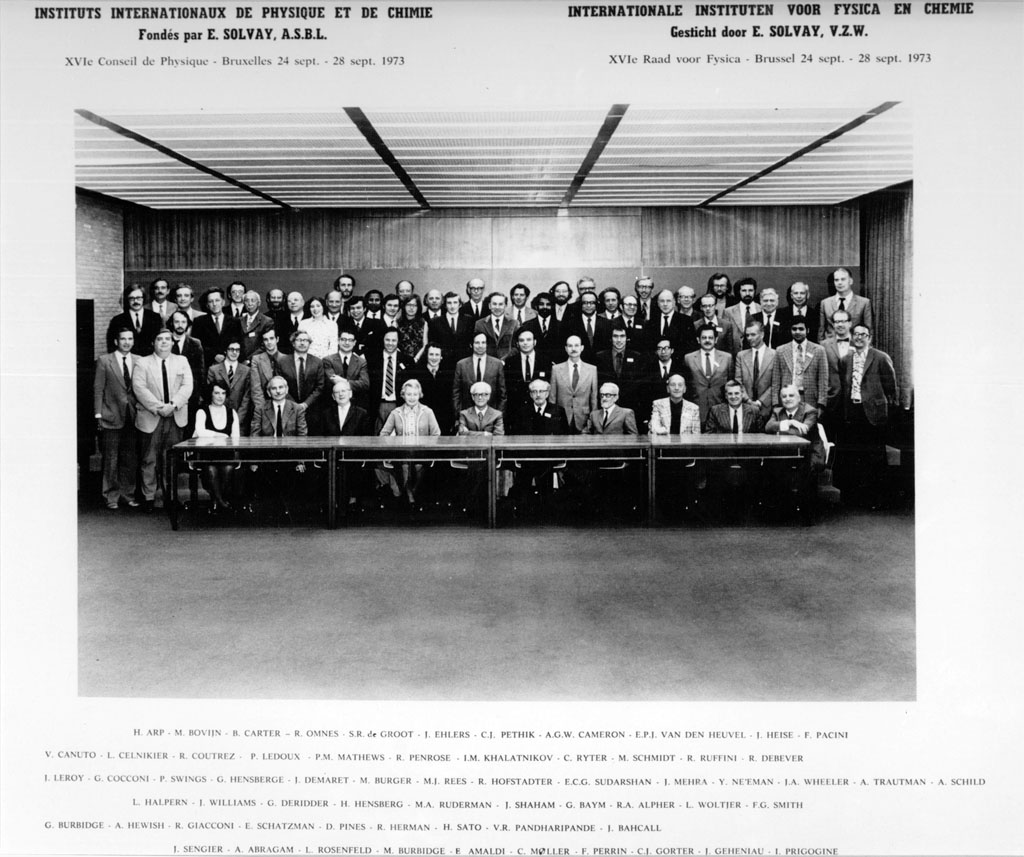}
	\label{solvay1973}
	\caption{Solvay meeting of 1973.}
\end{figure}
One very special occasion took place %bob happened 
in Moscow. One day I was visiting Yakov Borisovich in his Institute. He said ``Come and see a present I received from my friends in Minsk, where I was born.'' And he showed me a bronze statue of himself. I told him ``Congratulations, I can finally say that I have a friend with the bronze face!'' using the Italian meaning ``faccia di bronzo'' which are %bob
not very complementary words addressed to someone who is insensitive to problems.
Full of these memories I was delighted to see in the city of Minsk, now reconstructed and rebuilt, in the serenity of the spring his statue in form of a monument in front of the Academy of Sciences, see figure \ref{m}.

In 1985 I decided to create an international consortium dedicated to the field of relativistic astrophysics, the International Center for Relativistic Astrophysics (ICRA). This consortium relates the University of Rome ``La Sapienza'' to the University of Stanford, and the Space Telescope Institute at the USA, the University of Science and Technology in Hofei, China, the Specola Vaticana and the ICTP. It was coherently founded by George Coyne, Li-Zhi Fang, Francis Everitt, Riccardo Giacconi, Abdus Salam, and myself, see figure \ref{fig:Coyne}.

The most unique occasion with Zeldovich came in 1986 in Rome during the visit of the four delegations of the space research program of Europe, Japan, Soviet Union and the USA in occasion of the Halley comet mission. ICRA organized the meeting at ``La Sapienza'' and the Vatican. It was the first time Zeldovich could come to the West as a member of a very exceptional delegation created by Roald Sagdeev for this epochal meeting. There are many anecdotes with Zeldovich being shocked by a number of cars in the Italian streets and proposing to help himself with one since in his opinion it would be impossible to trace back the real owner. I did successfully convince him no to proceed in such an idea. Entering in the ``Sala Regia'' in the Vatican he attempted to seat in the first row and to my request to take his assigned seat in the 21st row seeing all the remaining ones still empty he said ``Nobody will notice me in the first row.'' I insisted that he should come back to the seat assigned to him by the Vatican ceremonial office. After few minutes he realized that the first rows were occupied on one side by the cardinals, the bishops and personnel of the Vatican, and on the other side by the ambassadors to the Vatican all in their sumptuous vests. Certainly the presence of Zeldovich in the first row would have been quite obvious and unjustifiable! But the surprises were not yet over. I was supposed to introduce him to the Pope during the audience with the members of the delegations. And I saw Zeldovich approaching with a clearly large object under his jacket. I was terrified, see figure \ref{p1}. 

Suddenly Zeldovich opened the jacket in front of John Paul II, extracted two books and put them %bob  gave then in 
into the hands of the Pope John Paul II, see figure \ref{p2}. His holiness said ``Thank you very much, professor Zeldovich'', and to this with a very loud voice which penetrated the entire ``Sala Regia'' Zeldovich forcefully replied ``Not just `thanks'! These are fifty years of my work!'' There was a great laugh from everybody as they relaxed. Later on John Paul II recalled that this was one of joyful audiences he had ever had. And he kept the two large red volumes over his white robe during the entire audience, see figure \ref{p3}.

Finally I would like to remark that a great scientist can even make a great discovery when he participates in some irrational actions. In the late fifties when the race to the Moon between the US and the Soviet Union was on someone proposed to show the great technical ability in the space vehicles and in the nuclear technology proposing to the Soviet superiority to explode at a fixed time an atomic bomb on the Moon\footnote{Different versions exist of this story. Some presented direct involvement of Zeldovich \cite{FM}, some show Zeldovich as an opponent of this idea on technical grounds \cite{GK}.}. This awful project fortunately was never implemented. Nevertheless it was one of the motivations to develop a highly secret mission from the United States in order to test the no proliferation agreement: the Vela satellites. These satellites were conceived to patrol all the region around the Earth and the Moon for possible nuclear explosions! Everybody knows today that this led to the discovery of gamma-ray bursts and we were very honored and pleased to announce their discovery at the 1972 AAAS meeting in San Francisco which was chaired by Herb Gursky and myself \cite{RGu}.

\begin{figure}[!ht]
	\centering
		\includegraphics[width=3in]{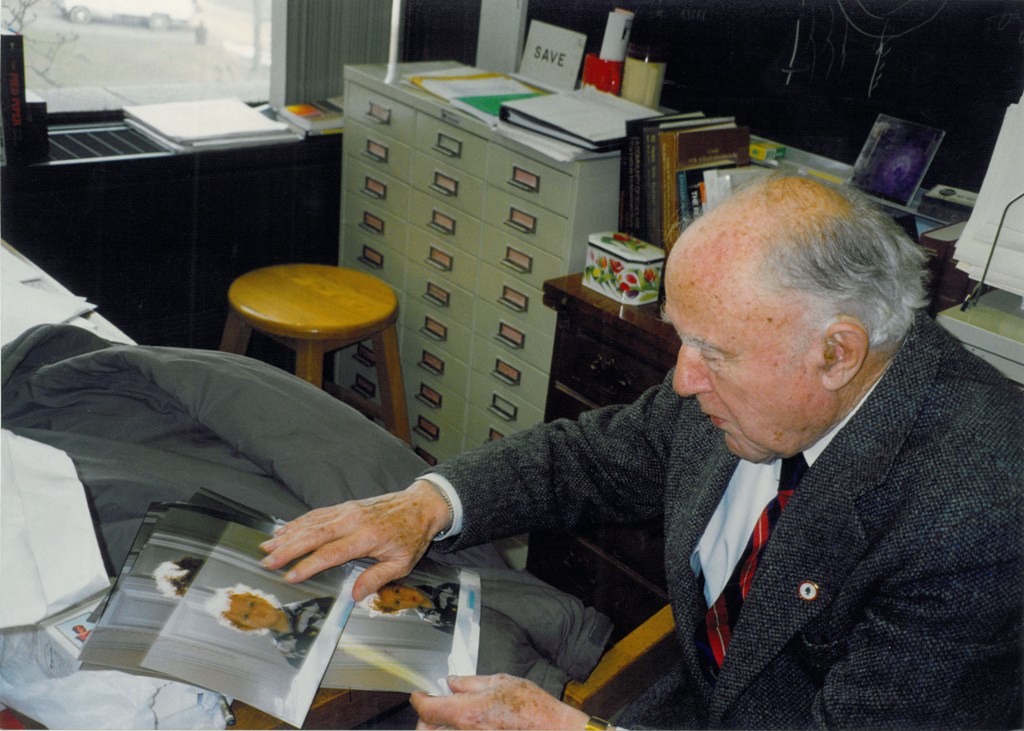}
	\label{Wheeler}
	\caption{Jonhy enjoying the pictures of Jacopo in 1999.}
\end{figure}
\begin{figure}[!ht]
	\centering
		\includegraphics[width=3in]{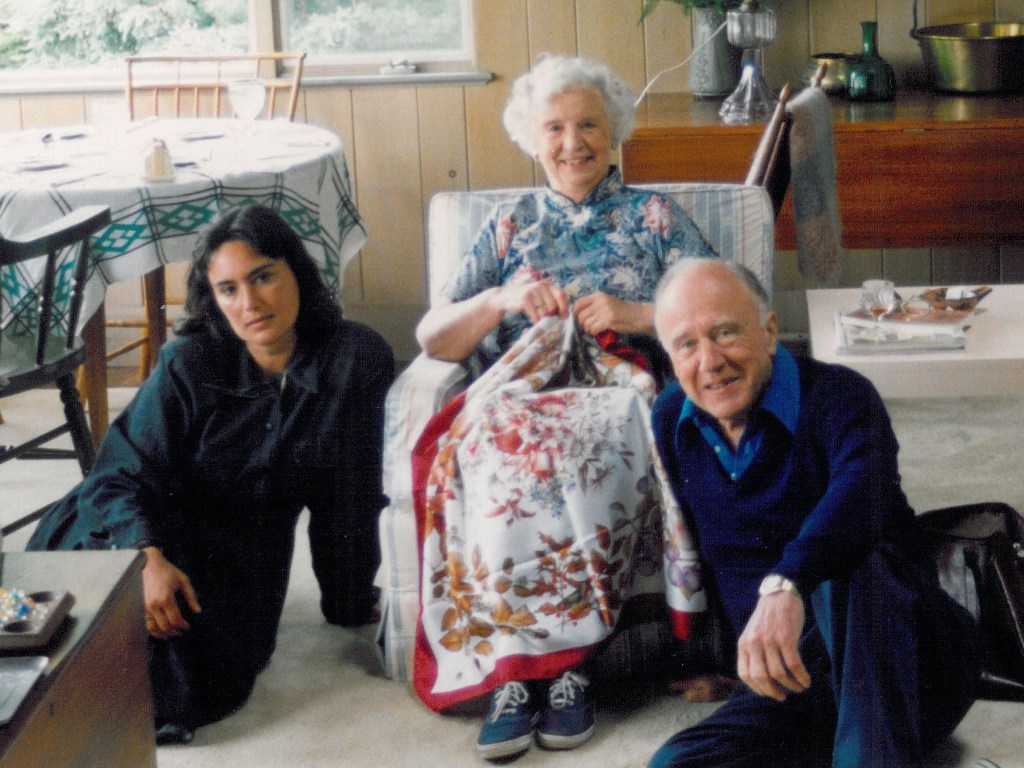}
	\caption{Picture of Ginette and Johny Wheeler with Anna in High Island with Ginette holding one of her preferred Gucci scarf.}
	\label{Wheeler2}
\end{figure}
\begin{figure}[!ht]
	\centering
		\includegraphics[width=3in]{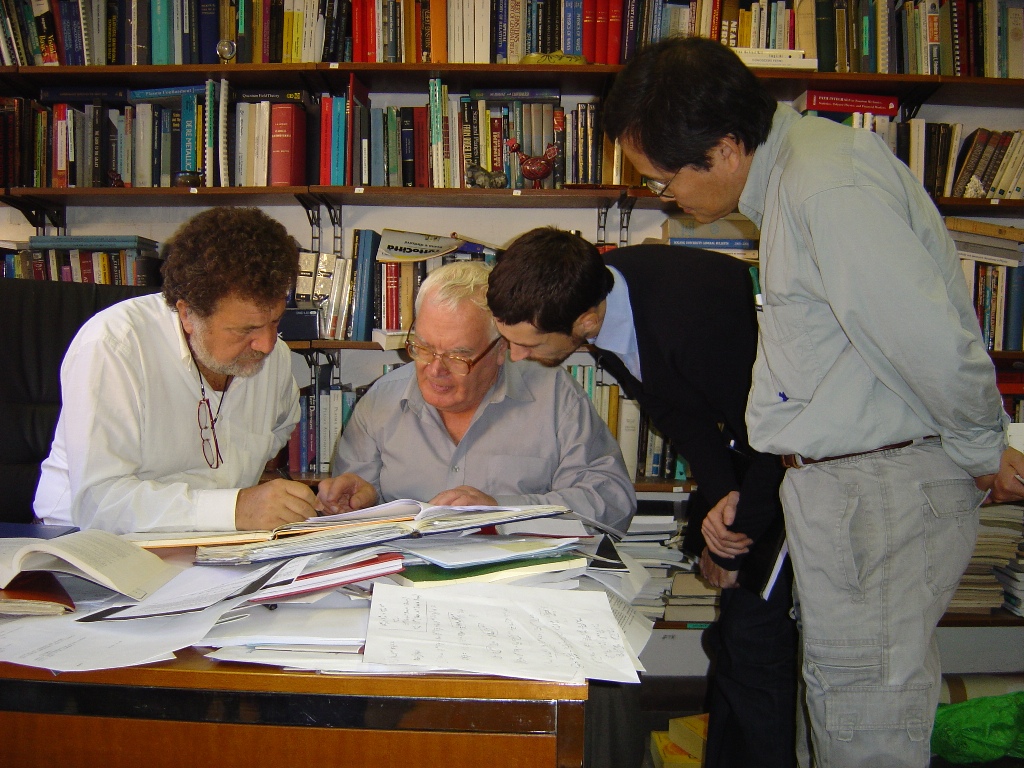}
	\caption{Picture taken in my office at ``La Sapienza'' of Vladimir Popov surrounded by Gregory Vereshchagin, She-Sheng Xue and myself in 2006.}
	\label{fig:pvx}
\end{figure}
\begin{figure}[!ht]
	\centering
		\includegraphics[width=5.9in]{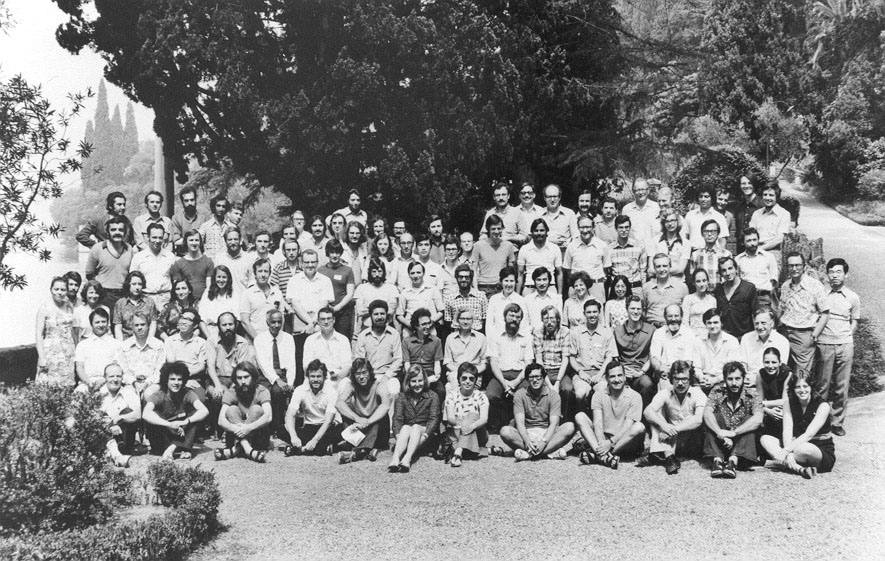}
	\caption{The picture of the participants of the Varenna summer school. In the second row Anthony Hewish (Nobel Prize, 1974), Joe Taylor (Nobel Prize, 1993), Subrahmanyan Chandrasekhar (Nobel Prize, 1983) and Riccardo Giacconi (Nobel Prize, 2002).}
	\label{fig:Varenna}
\end{figure}
\begin{figure}[!ht]
	\centering
		\includegraphics[width=2.6in]{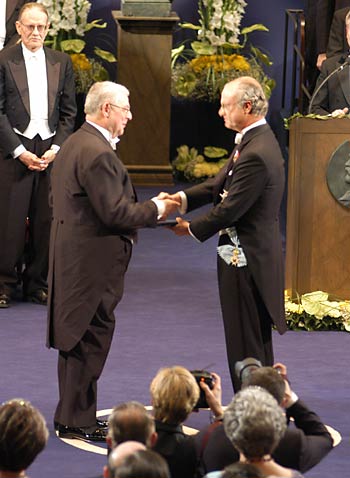}
	\caption{Picture of Riccardo Giacconi receiving the Nobel Prize.}
	\label{fig:Giacconi}
\end{figure}
\begin{figure}[!ht]
	\centering
		\includegraphics[width=3in]{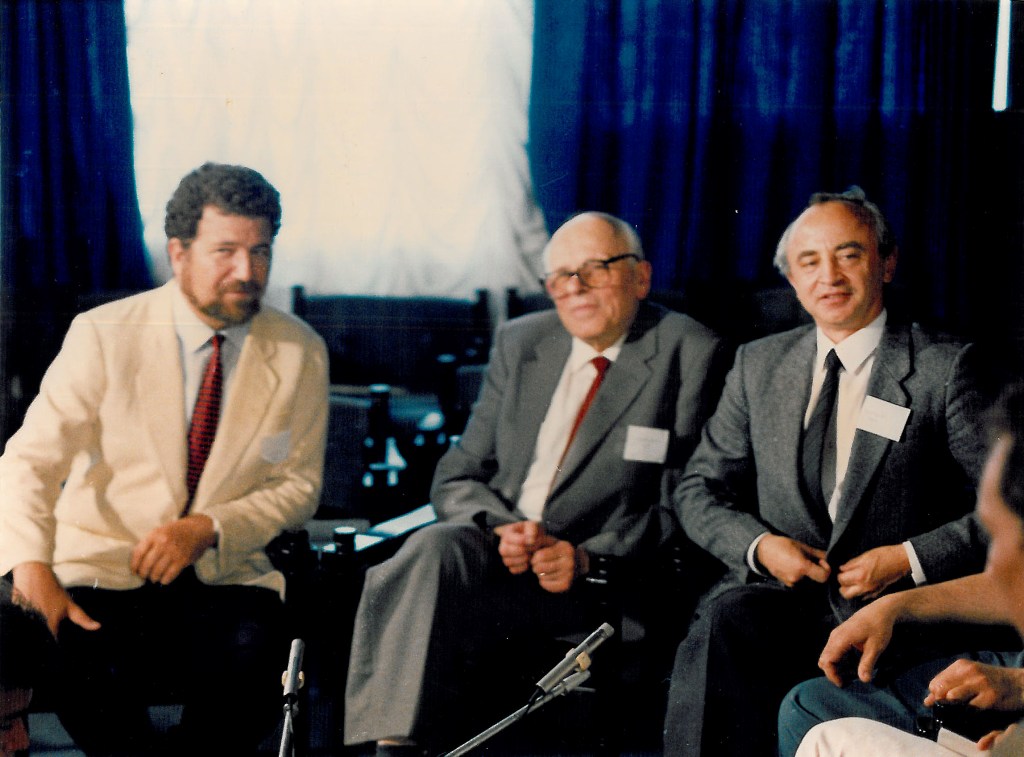}
	\label{fig:rsn}
	\caption{Television broadcast made by Igor Novikov, Andrei Sakharov and myself in the celebration of Alexander Alexandrovich Friedman's 100th Anniversary, Leningrad, 1988.}
\end{figure}
In 1987 I visited Zeldovich in Moscow for the last time. There was a meeting at the Academy of Sciences on cosmology. While he went to deliver his talk he asked me to keep his jacket with the three gold stars and red stripes of the Hero of Socialist Labor. He was among the few people to have three such decorations. They told me that even Stalin had only one such ``star''. I was not surprised. By that time I had become aware of his many contributions in ignition, combustion, explosions as well as of his work with Yulii Khariton and Igor Kurchatov on the atomic bomb. Slowly but inevitable I became also aware of the role of John Wheeler in the American H-bomb project. Of course it was clear they had done an enormous work in the physics of the bomb and also it was evident that they had learned one of the greatest amount of physics reachable at the time. 

When it came to the work on Relativistic Astrophysics I was surprised to see that this vast quantity of knowledge in physics they had acquired in making the bombs did not help as much as one would have expected. They were somewhat overshooting and did not catch the beauty, the different and possibly more profound physical scientific complexity, and also the conceptual simplicity of the new phenomena. In the case of Wheeler the interactions with him during the first years in Princeton had be tremendously intense. At times we were working 13 hours a day. We wrote that celebrated article for Phyiscs Today \cite{RW}, recently reprinted \cite{RW2}, in which we were presenting for the first time a Black Hole as a physical object and not just as a mathematical solutions. Such an object was indeed interacting actively with the rest of the Universe by a vast amount of energy, in principle extractable: the rotational and the electromagnetic energy. These works were received an exponential growth with the coming to Princeton of Demetrios Christodoulou from Greece at the age of 16. When he started his thesis of PhD at the age of 18 Demetrios approached the problem suggested by Wheeler of the collapse of a scalar field forming a black hole which he finally solved in 2009 \cite{CMG12}. A second part of his thesis was developed under my guidance \cite{Kerr} which has led to the general mass formula of the black hole \cite{RC}, see figure \ref{fireplace}. Interestingly precisely these concepts have made later the Black Holes through their ``Blackholic energy'' the explanation of Gamma Ray Bursts \cite{DR}: the largest instantaneous  energy sources in the Universe second only to the Big Bang \cite{Kerr,Brazil,Report}. In collaboration with Rees we also wrote a book giving guidelines for the study of Black Holes, Gravitational Waves and Cosmology \cite{RRW}. The field of Relativistic Astrophysics started to grow exponentially after the introduction of X Ray Astronomy by Riccardo Giacconi and his group \cite{GRS}. Paradoxically Wheeler interest started to depart from these topics and drifted toward a (possibly too) vast field of exploring the world of mathematics in the quest for better expressing the laws of physics, see also my recollections in \cite{Kerr}. It was that time in which I proposed the paradigm for the first identification of a Black Hole in our Galaxy \cite{Leach}, see figure \ref{prize}.

A profound separation of scientific interests had already occurred in those days at the Les Houches summer school: the first one solely dedicated to black holes \cite{LH}. After that event I dedicated myself to the study of Black Holes larger than 3.2 solar masses. While S. Hawking and his group directed all the attention to mini black holes (see e.g. \cite{RK}). The field of matter accretion on a Black Hole was not developed in the West and became dominated by the Russian (see Titarchuk contribution to this volume) and Indian schools (see Chakrabarti contribution to this volume). In the case of Wheeler a different point of view on the role of European scientists in the United States of America emerged, and a separation of our scientific interest became manifest in the 1973 Solvay meeting (see figure \ref{solvay1973}), which was followed by my return to Europe. These differences did not affect in any way the deep friendship between us extended to our families, see figures \ref{Wheeler} and \ref{Wheeler2}.
 
In the case of Zeldovich some similar event happened. I was trying to make him appreciate the beauty of the work I was developing with an American hero of Relativistic Astrophysics, Jim Wilson, himself a distinguished participants of the American Bomb projects. The work on the relativistic magnetohydrodynamics effect around Black Holes have today reached the greatest interest for microquasars and active galactic nuclei explanations \cite{Punsly}. To that he was answering with his interests toward the possible radiation of a rotating sphere due to quantum effects. To me that work did nor appear so promising in view of the intrinsic stability imposed by quantum effects on a rotating system.

Thinking over my scientific discussions with Zeldovich I was especially admiring his work with Vladimir Popov on heavy nuclei, as expressed in our recent report \cite{Report}. On this topic see also Popov's contribution in this book. This topic has become central to our current research, see figure \ref{fig:pvx}.

In all my discussions with Zeldovich through the seventies I was particularly eager to illustrate to him my work on the black hole identification and to observe his feedback. Much of these works, following the Solvay meeting, were summarized in our celebrated Varenna summer school, see figure \ref{fig:Varenna}. This basic work then appeared in the book \cite{RGi} which is currently being reprinted \cite{RGinew}. That epochal meeting in the scientific content was followed until today by three Nobel Prize winners among the lecturers as S. Chandrasekhar (1983), J. Taylor (1993), and R. Giacconi (2002), see figure \ref{fig:Giacconi}.

But let us return after this digression to my last meeting with Zeldovich. While he was speaking Sakharov entered the room and sat in the first row near me. He had just been permitted to return to Moscow after the Gorky exile. I had just been helping at the University of Rome to attribute to him a \emph{laurea honoris causa - in absenzia}. I looked at him closely: the face had changed from the Tbilisi days, his smile was gone and his gentle aspect had been modified. Even the structure of the face was somewhat more tense with a more prominent jaw. I gave my hand to him: ``Ruffini, Italy'' and his immediate answer recalling a serene expression resembling the old days ``Sakharov, Soviet Union!''

In June 1988 on the hundredth anniversary of the birth of Alexander Alexandrovich Friedman we went to Leningrad with Werner Israel and a few other relativists. It was a very emotional occasion to find the tomb of Friedman and put some flowers on it. Yakov Borisovich Zeldovich had died on December 2, 1987. This was the occasion of a trip by night sleeping train between Moscow and Leningrad with my wife Anna. The next compartment on that train was occupied by Andrei Sakharov and Elena Bonner. The day after a memorable broadcast from the television was made by Igor Novikov, Andrei Sakharov and myself in the celebration of Alexander Alexandrovich Friedman, see figure \ref{fig:rsn}.

\end{document}